\documentclass[aps,prl,showpacs,amsmath,twocolumn,amssymb,superscriptaddress,letterpaper]{revtex4}
\usepackage{times}
\usepackage{amsfonts}
\usepackage{mathrsfs}
\usepackage{graphicx}
\usepackage{dcolumn}
\usepackage{bm}
\usepackage{color}

\usepackage[colorlinks,bookmarks=false,citecolor=blue,linkcolor=red,urlcolor=blue]{hyperref}
\usepackage{color}
\usepackage{siunitx}
\usepackage{bibunits}

\def\be{\begin{equation}}       \def\ee{\end{equation}}
\def\bea{\begin{eqnarray}}      \def\eea{\end{eqnarray}}

\begin{document}
\begin{bibunit}
\title{ CaFeAs$_2$: a Staggered Intercalation of Quantum Spin Hall and High Temperature Superconductivity}

\author{Xianxin Wu}
\affiliation{ Institute of Physics, Chinese Academy of Sciences,
Beijing 100190, China}

\author{Shengshan Qin}
\affiliation{ Institute of Physics, Chinese Academy of Sciences,
Beijing 100190, China}

\author{Yi Liang}
\affiliation{ Institute of Physics, Chinese Academy of Sciences,
Beijing 100190, China}

\author{Congcong Le}
\affiliation{ Institute of Physics, Chinese Academy of Sciences,
Beijing 100190, China}

\author{Heng Fan }  \affiliation{ Institute of Physics, Chinese Academy of Sciences,
Beijing 100190, China}
\affiliation{Collaborative Innovation Center of Quantum Matter, Beijing, China}

\author{Jiangping Hu  }\email{jphu@iphy.ac.cn} \affiliation{ Institute of Physics, Chinese Academy of Sciences,
Beijing 100190, China}\affiliation{Department of Physics, Purdue University, West Lafayette, Indiana 47907, USA}
\affiliation{Collaborative Innovation Center of Quantum Matter, Beijing, China}

\date{\today}

\begin{abstract}

 We predict that  CaFeAs$_2$,  a newly discovered iron-based high temperature (T$_c$) superconductor, is a staggered intercalation compound that integrates topological quantum spin hall (QSH) and superconductivity (SC). CaFeAs$_2$ has a structure with staggered CaAs  and FeAs layers. While the FeAs layers are known to be responsible for high T$_c$  superconductivity,   we show that with spin orbital coupling each CaAs layer is a   $Z_{2}$ topologically nontrivial two-dimensional QSH insulator and the bulk is a  3-dimensional weak topological insulator.   In the superconducting state,  the edge states in the CaAs layer  are natural 1D topological superconductors.  The staggered intercalation of QSH and SC provides us an unique opportunity to realize and explore novel physics, such as  Majorana modes and Majorana Fermions chains.

\end{abstract}


\maketitle
Topological insulator(TI) materials\cite{Kane2005,Hasan2010,Qi2011} and high T$_c$ superconductors\cite{Norman2003,Kamihara2008,Johnston2010}  are two types of intriguing materials with both important fundamental physics  and potential revolutionary applications\cite{Nagaosa2007,Seradjeh2009,Essin2009,Kamihara2008}.  Both materials are featured with a bulk gap in  the single electron excitation spectrum. In the former,  the bulk gap    protects  surface or edge states depending on dimensionality,  which are caused by the topological character change from the inner to outer materials. In the latter, the bulk gap protects  the Cooper pairs formed by two electrons, thus the superconductivity.

 Recently, there has been a very fascinating prediction of  the existence of Majorana fermions, arising as a quasi-particle excitation\cite{Fu2010,Linder2010} when  the superconductivity  is  induced in a topological insulator.   Besides their fundamental physical significance, Majorana modes can play a crucial role in topological quantum computation because of the topologically protected qubit and a possible realization of non-Abelian braiding\cite{Kitaev2003}.
Consequently, there have been a wealth of proposals for the experimental realization of Majorana fermions. The feasible systems include quantum wires in proximity to ordinary superconductors\cite{Oreg2010}, semiconductor-superconductor heterostructure\cite{Lutchyn2010,Mourik2012}, TIs coupled to superconductors\cite{Fu2010,Linder2010} and Al-InAs nanowire topological superconductors\cite{Das2012}.
However, in all these proposals, the high temperature superconductors have never been candidates in this  integration process with  topological insulators  because of their extreme short coherent length and structural  incompatibility.

Can we integrate topological insulators together with high T$_c$  superconductors or  is there a  material that naturally integrates both?  Compared with conventional superconductors, both iron-based\cite{Kamihara2008,Ren2008,Wang2008} and Copper-based high T$_c$ superconductors have  much higher T$_c$,  much larger superconducting gaps and  much higher upper critical magnetic fields.  Therefore, a positive answer to this question can allow future devices that utilize both topology and SC  to operate  not only at much higher temperature, but also much more robustly.

\begin{figure}[t]
\centerline{\includegraphics[height=7 cm]{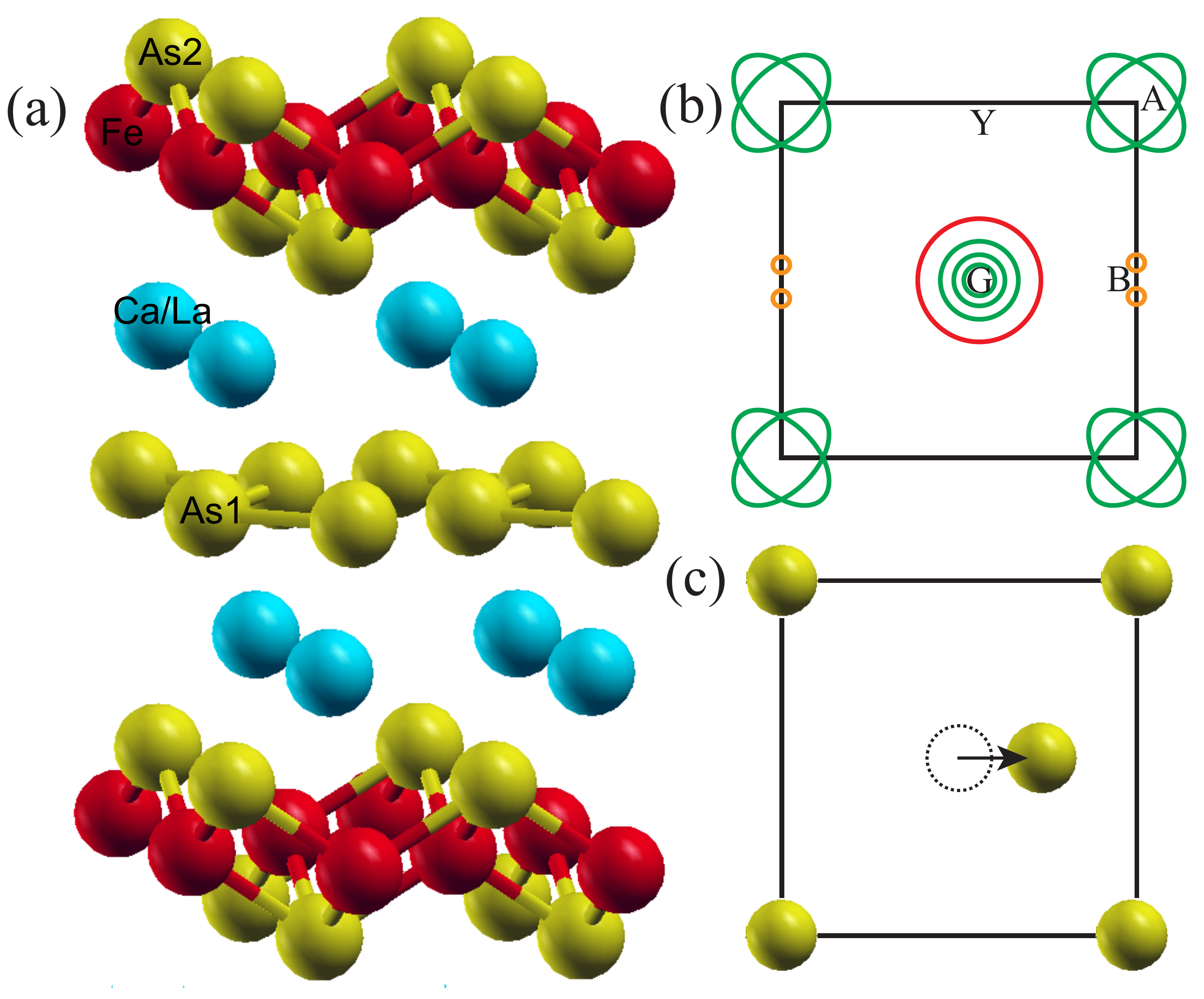}}
\caption{(color online). (a) Schematic view of the crystal structure of Ca$_{1-x}$La$_x$FeAs$_2$. (b) Fermi surfaces(FSs) for CaFeAs$_2$. The green circles and ellipses FSs are contributed by the FeAs layers. The red circle FS is attributed to $p_z$ orbitals of As-1 and $d$ orbitals of Ca. The orange circles FSs are attributed to $p_x$ and $p_y$ orbitals of As-1. (c) Lattice model for the As layer in CaFeAs$_2$.
 \label{lattice} }
\end{figure}

Here we  show that such an material has already existed in   iron-based high T$_c$   superconductors\cite{Kamihara2008}. The material is the recently synthesized  (Ca,Pr)FeAs$_2$\cite{Yakita2013} or Ca$_{1-x}$La$_x$FeAs$_2$\cite{Katayama2013} with $T_c$ over 40 K. The (Ca,Pr)FeAs$_2$ has a structure with a staggered intercalation between the chain-like As layers  and Fe-As layers along c-axis as shown in Fig.\ref{lattice}(a).  The Fe-As layers, as the common components in all iron-pnictide superconductors, are known to be responsible for high T$_c$ superconductivity.  Previously, it was also shown that the chain-like As layers provide an additional band structure that  is absent in any other iron-pnictide materials\cite{Wu2014}: the $p_x$ and $p_y$ orbitals of the As atoms in the chain-like layers provide an anisotropic Dirac Cone near the Fermi level.

Here we predict that each chain-like As layer is a topological insulator. Thus we predict that CaFeAs$_2$ is a staggered intercalation compound that integrates both QSH and SC and is an ideal system for the realization of Majorana related physics.  In the following, we will show that  the Dirac cone attributed to the chain-like layers  is  gapped by the spin orbital coupling(SOC) with a gap over 100 $meV$. More importantly, we  demonstrate that it is topological nontrivial.   Thus, below $T_c$, we have a natural topological insulator-superconductor hybrid structure in Ca$_{1-x}$La$_x$FeAs$_2$. The topological protected edge state from the chain-like layer   becomes an one-dimensional  topological superconductor.  With  stacking along c-axis, it becomes a quasi one-dimensional topological superconducting system which can be manipulated to  realize novel topological related physics.

We start  a brief discussion of  the crystal and band structures of CaFeAs$_2$ which have been detailedly discussed \cite{Wu2014}.
Fig.\ref{lattice}(a) shows the crystal structure of CaFeAs$_2$, which contains alternately stacked FeAs and CaAs ( or As chain-like) layers.  It has been shown in\cite{Wu2014}  that  the electronic band structures near Fermi surfaces can be divided into weakly coupled three parts including the d-orbital bands from Fe-As layers,  the bands from  the $p_x$ and $p_y$ orbitals of As-1 and the bands from the $p_z$ orbitals of As-1. Fig.\ref{lattice}(b) shows the Fermi surfaces(FSs) and Fig.\ref{lattice}(c)  shows the displacement of the As atoms in the CaAs layers.  In Fig.\ref{lattice} (b), the normal FSs in green are contributed by  the FeAs layers and the small orange FSs are attributed to $p_x$ and $p_y$ orbitals of As-1\cite{Wu2014}.  The bands which contribute to these FSs are very two-dimensional and have little c-axis dispersion. The red FS is attributed to $p_z$ orbitals of As-1 and $d$ orbitals of Ca. The couplings among these three different band structures are very weak. This is consistent with the fact that the bands from the d-orbitals of Fe in FeAs layers  in all other iron-pnicitides  are almost identical and are insensitive to atoms outside the layers.   Because of the displacement, the As atoms in  the CaAs layers form zigzag chains with a distorted checkerboard lattice in which an As atom moves in the $x$ direction and is off the center position, as shown in Fig.\ref{lattice} (c).  This movement breaks the $S_4$ symmetry at Fe sites.  In the momentum space, lowering the symmetry causes the degeneracy lift on the band structure  in the $k_x=\pi$ plane.

\begin{figure}[t]
\centerline{\includegraphics[height=4.5 cm]{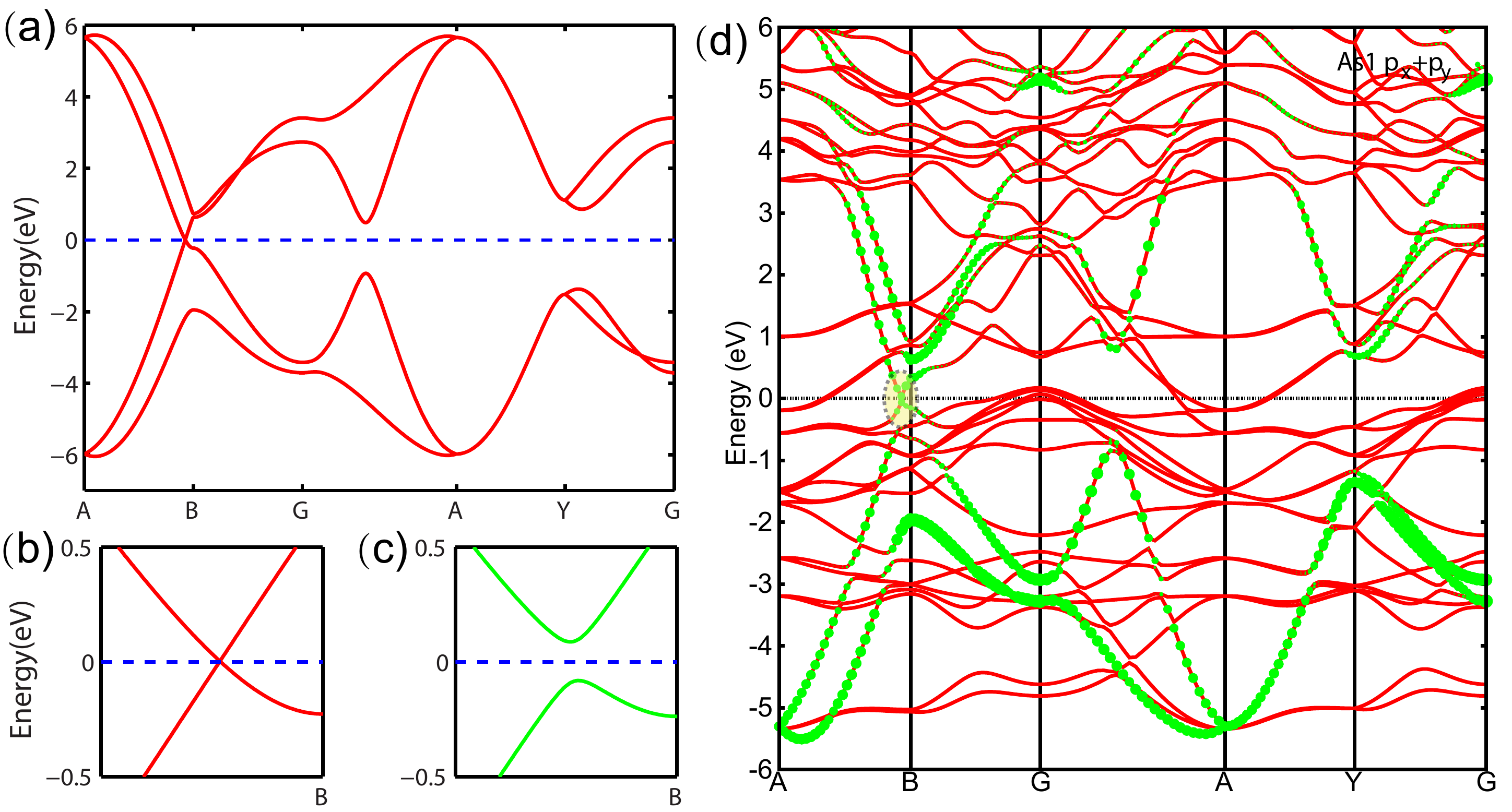}}
\caption{(color online). Band structures for As layers and CaFeAs$_2$. (a) Band structure without SOC. The zoom-in band structure near the Fermi level  without SOC (b) and with SOC (c). The strength of SOC is 0.19 eV. (d) The LDA band structure of CaFeAs$_2$ with SOC. The Dirac cone is gapped by SOC and the gap is about 0.1 eV. The sizes of green circles are proportional the weights of $p_x$ and $p_y$ orbitals.
 \label{band} }
\end{figure}

 A two-dimensional four-band model has been derived  to capture  the band structure attributed to $p_x$ and $p_y$ orbitals of two As-1 \cite{Wu2014} in a unit cell. As a unit cell includes  two As-1 atoms, we divide the As-1 lattices  into two sublattices.  We introduce the operator $\phi^\dag_{\textbf{k}\sigma}=[c^\dag_{ax\sigma}(\textbf{k}),c^\dag_{ay\sigma}(\textbf{k}),c^\dag_{bx\sigma}(\textbf{k}),c^\dag_{by\sigma}(\textbf{k})]$, where $c^\dag_{\alpha \beta\sigma}(\textbf{k})$ is a Fermionic creation operator with  $\sigma$,  $\beta$ and $\alpha$ being spin, orbital and sublattice indices respectively. The tight-binding Hamiltonian can be written as:
 \begin{eqnarray}
 H_{TB}=\sum_{\textbf{k}\sigma}\phi^\dag_{\textbf{k}\sigma}h(\textbf{k})\phi_{\textbf{k}\sigma}.
 \end{eqnarray}
  The matrix elements in the Hermitian $h(\textbf{k})$ matrix are given by
\begin{eqnarray}
\label{caas_tb}
&&h_{11}=h_{33}=\epsilon_X+2t^{11}_{1}cosk_x +2t^{11}_{2}cosk_y, \nonumber\\
&&h_{13}=(2t^{13}_{1}e^{i(x_0-1)k_x}+2t^{13}_{2}e^{ix_0k_x})cos(k_y/2),\nonumber\\
&&h_{14}=h_{23}=-(2it^{14}_1e^{i(x_0-1)k_x}+2it^{14}_2e^{ix_0k_x})sin(k_y/2),\nonumber\\
&&h_{22}=h_{44}=\epsilon_Y+2t^{11}_{2}cosk_x +2t^{11}_{1}cosk_y,\nonumber\\
&&h_{24}=(2t^{24}_1e^{i(x_0-1)k_x}+2t^{24}_2e^{ix_0k_x})cos(k_y/2),
\end{eqnarray}
with $x_0$ being the difference between the $x$ components of $a$ and $b$ sublattices.  The corresponding tight binding parameters are specified in unit of $eV$ as,
\begin{eqnarray}
&&\epsilon_X=-0.30, \quad \epsilon_Y=-0.109, \quad t^{11}_1=-0.149,  \nonumber \\
&&t^{11}_2=0.128, \quad t^{13}_1=0.89, \quad t^{13}_2=0.649, \nonumber \\
&&t^{14}_1=1.169,\quad t^{14}_2=-1.740, \quad t^{24}_1=0.567  \nonumber \\
&& t^{24}_2=1.213.
\label{hopping}
\end{eqnarray}
The band structure of the model is shown in Fig.\ref{band}(a) and there is an anisotropic Dirac cone near B, shown in Fig.\ref{band}(b), but not near Y because the $S_4$ rotation symmetry is broken in this system. It is important to note that this Dirac point is very close to  the Fermi level.   The band degeneracy in A-Y line is protected by a hidden symmetry\cite{Wu2014}.

Now we consider the SOC  term in the As-1 atoms.   The SOC  can be written as
\begin{eqnarray}
H_{so}=i\lambda/2\sum_{\alpha \sigma\textbf{k}}\sigma c^{\dag}_{\alpha x\sigma}(\textbf{k}) c_{\alpha y\sigma}(\textbf{k})+h.c.
\end{eqnarray}
For a  general purpose, we can also add a term that describes a stagger sublattice potential given by
\begin{eqnarray}
H_{p}=\lambda_{\nu}/2\sum_{\beta\sigma\textbf{k}}(c^{\dag}_{a\beta\sigma}(\textbf{k}) c_{a\beta\sigma}(\textbf{k})-c^{\dag}_{b\beta\sigma}(\textbf{k}) c_{b\beta\sigma}(\textbf{k})).
\end{eqnarray}  This term vanishes for  the CaAs layers in a bulk CaFeAs$_2$.   However, in principle, the term can be generated if the As layer is a surface layer or by additional lattice distortions caused by external forces.  The SOC strength  $\lambda$ of As is about 0.19 eV\cite{Wittel1974} and $\lambda_{\nu}$ can be considered as an adjustable parameters.

 The full Hamiltonian that describes the band structure of  the CaAs layer is given by
 \begin{eqnarray}
 H=H_{TB}+H_{so}+H_{p}.
 \label{fullh}
 \end{eqnarray}  We first show that  this Hamiltonian produces the band structure calculated by the standard LDA method when the SOC is considered.  With SOC, the LDA band structure of CaFeAs$_2$ is shown in Fig.~\ref{band}(d) and the gap(shaded ellipse in Fig.\ref{band}(d)) of CaAs layers is about 0.1 $eV$. Taking $\lambda_{\nu}=0$ and $\lambda = 0.19 eV$ in  Eq.\ref{fullh}, the Hamiltonian produces a similar band structure  to the LDA result as shown in Fig.\ref{band}(a,b,c) which show the major difference between the two cases. With $\lambda = 0.19 eV$, the gap at Dirac cone from  Eq.\ref{fullh} is about 0.17 $eV$  which is slightly larger than the LDA result. This result is expected because the SOC strength should be adjusted in a  tight-binding model.

\begin{figure}[t]
\centerline{\includegraphics[height=7 cm]{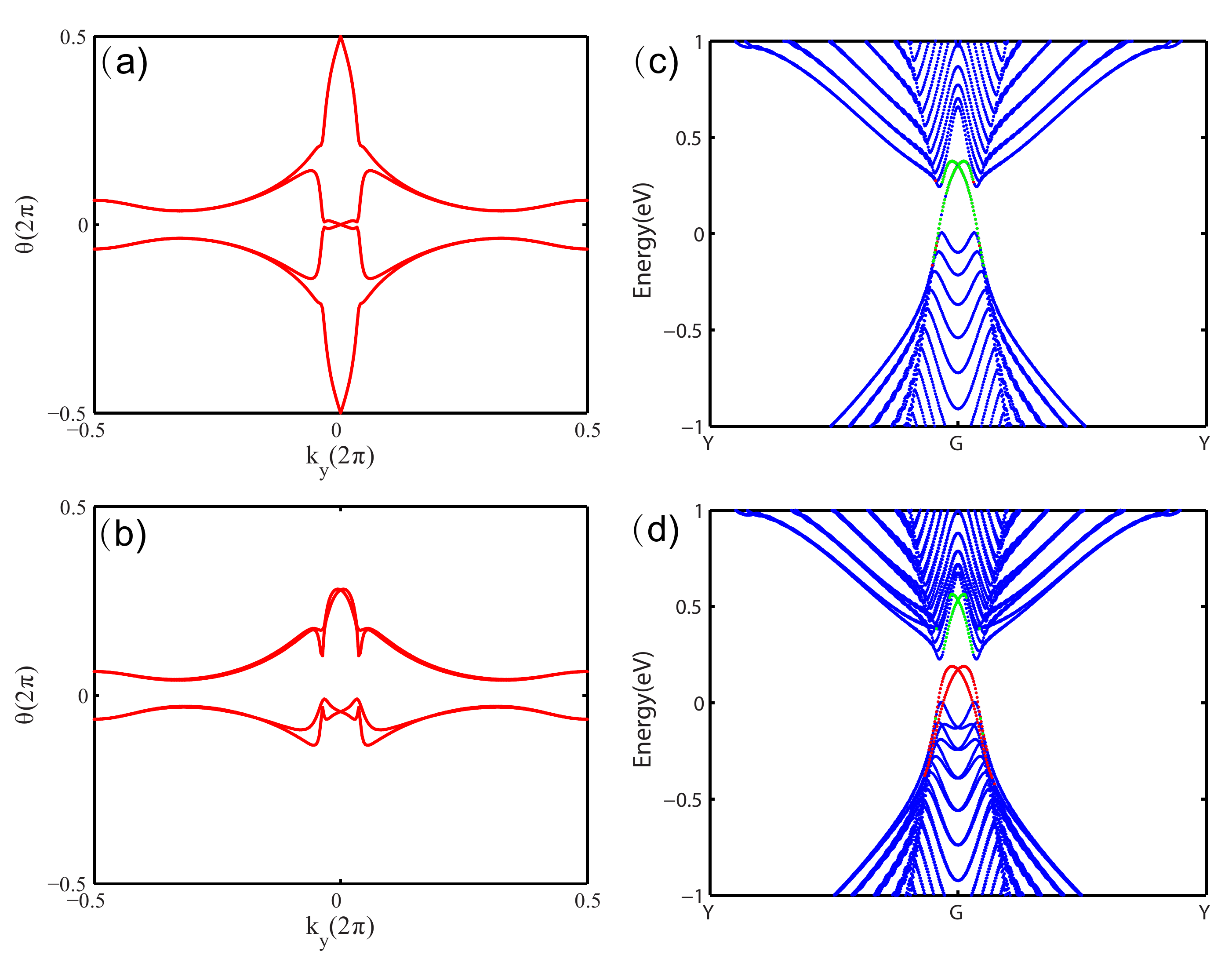}}
\caption{(color online). The evolution of Wannier function centers( (a) and (b)) and edge states( (c) and (d)) for our tight binding model. (a), (c) $\lambda=0.19$ eV and $\lambda_{\nu}=0$. The vertical axis shows the Wannier centers of the four occupied bands. (b), (d) $\lambda=0.19$ eV and $\lambda_{\nu}=0.19$ eV. The red and green lines denote edge states from different edges.
 \label{wannier_edge} }
\end{figure}

Is the above band topologically trivial or nontrivial?  To answer this question, we calculate the Wannier function centers of the occupied bands\cite{Yu2011}. With SOC, the evolution pattern of the Wannier centers and edge centers are given in
 Fig.\ref{wannier_edge}(a) and (c) where it is explicitly shown that the Wannier states switch partners  and there are edge states in the gap, a signature proving that the states are topologically nontrivial. The edge states are mainly contributed by $p_x$ orbitals.  Considering the crossing points between the evolution line and an arbitrary horizontal line in the half Brillouin zone, one can aslo find that  the number of the crossing points is odd, indicating that the Z$_2$ topological number is odd.  As a comparison,  we plot a  typical evolution pattern of the Wannier centers and edge states  in Fig.\ref{wannier_edge}(b) and (d) for a topologically trivial case.

\begin{figure}[t]
\centerline{\includegraphics[height=6.5 cm]{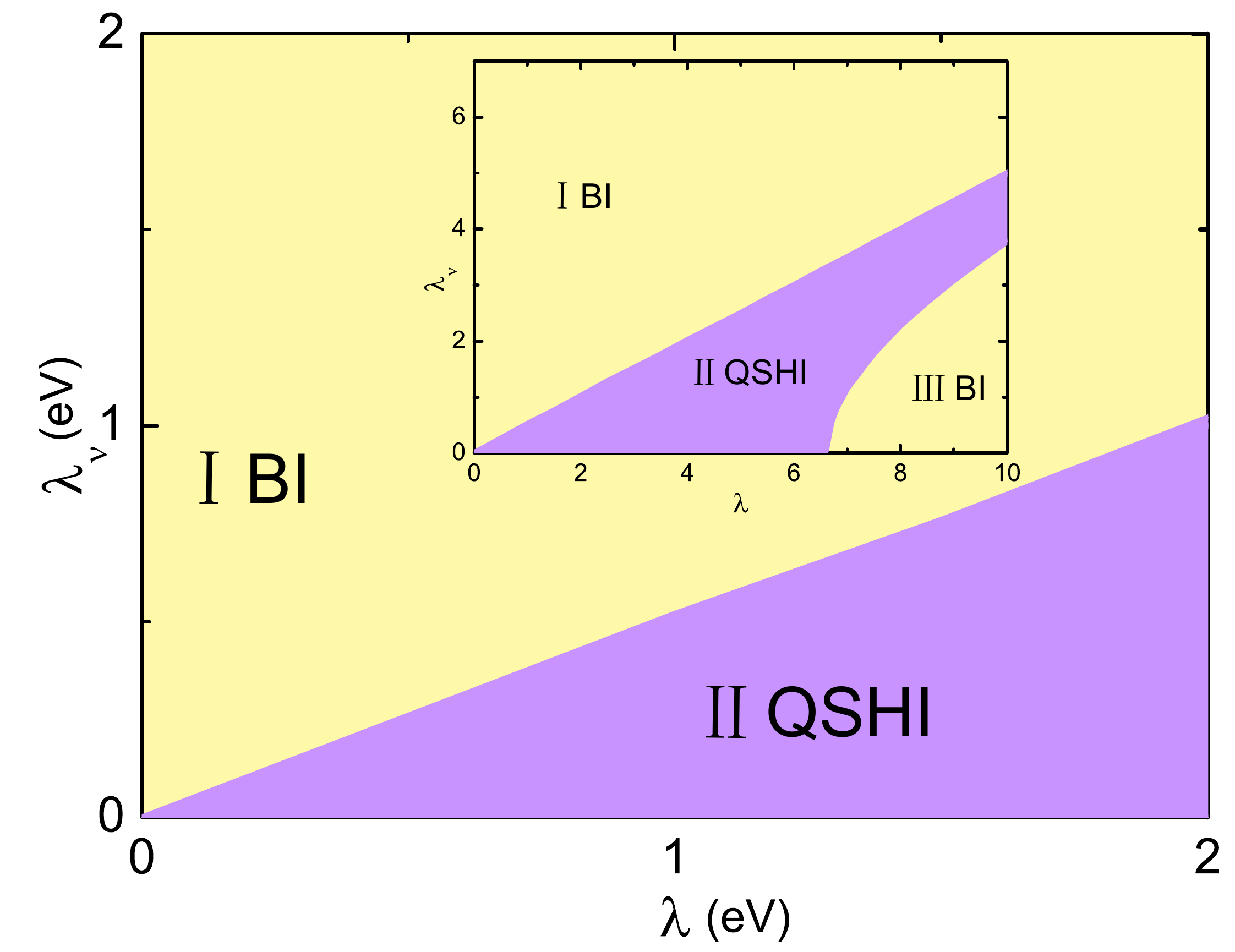}}
\caption{(color online). The topological phase diagram in $\lambda-\lambda_{\nu}$ plane for possible  material systems. The inset shows the phase diagram in the whole parameter space.
 \label{phase} }
\end{figure}

We can also obtain the phase diagram in  the $\lambda-\lambda_{\nu}$ plane for the model.  As shown in Fig.\ref{phase}, the topological phase transition from a topological trivial  band insulator(BI)  to a topological nontrivial QSH insulator takes place at the line $\lambda_{\nu}=\lambda/2$.  When  $\lambda_{\nu}>\lambda/2$, the system is a BI. The system becomes a topological insulator when $2\lambda_{\nu}<\lambda$.  We note that  for an extremely large $\lambda>6.5$ eV  which is larger than the band width, the system becomes a BI again.

In the above calculation, we omit the As-1 $p_z$ orbitals, which couple strongly with Ca $d$ orbitals and are well separated from the $p_x$ and $p_y$ orbitals in the band structure. Although they can couple  to $p_x$ and $p_y$ orbitals through SOC, the strength is  too weak to have any changes on above results as  the QSH insulator state for the real CaAs layer is rather robust  with a  bulk gap over 100 $meV$.  We have also calculated the local density of states of free standing As layers and the complete CaFeAs$_2$ system (see supplementary material \cite{supplemantary}). In both cases, the nontrivial edge states attributed to the As layers are very similar to those in Fig.\ref{wannier_edge}(b).  Furthermore,  the calculation shows that the bulk material CaFeAs$_2$ is a 3-dimensional weak topological insulator with a 3-dimensional topological invariance $Z_2=0;(001)$\cite{Fu2007}( see supplementary material \cite{supplemantary}).


In the QSH insulator phase, the edge states   are fully spin polarized and can be modeled by the Hamiltonian,
\begin{eqnarray}
H_{TI}=\sum_{\textbf{k}}\psi^{\dag}(\textbf{k})(vk\sigma_z-\mu)\psi(\textbf{k})
\end{eqnarray}
where $v$ is the edge-state velocity, $\mu$ is the chemical potential and $\psi^{\dag}_{\sigma}(\textbf{k})$ creates an electron with spin $\sigma$ and wave vector $\textbf{k}$ in the edge state bands.  Below T$_c$, the FeAs layer become superconducting. As the pairing in the FeAs is a full gapped s-wave spin singlet pair\cite{Johnston2010,Hirschfeld2011},   the superconducting proximity effect on the edge of As layers can be modeled with a Hamiltonian: \begin{eqnarray}
H_{S}=\sum_{\textbf{k}}\Delta\psi_{\downarrow}(-\textbf{k})\psi_{\uparrow}(\textbf{k})+h.c..
\end{eqnarray}
Moreover, for a general purpose, we can also introduce a Zeeman field which cants the spins away from the $z$ direction,
\begin{eqnarray}
H_Z=-h\sum_{\textbf{k}}\psi^{\dag}(\textbf{k})\sigma_x\psi(\textbf{k}).
\end{eqnarray}
Therefore, the full Hamiltonian for the edge states is given by,
\begin{eqnarray}
H_{edge}=H_{TI}+H_Z + H_{S}
\end{eqnarray}
  Specifically, when $\Delta=0$, the energy spectrum is $\epsilon_{\pm}(k)=-\mu \pm \sqrt{(vk)^2+h^2}$ and $\psi_{k\pm}$ create electrons with energy $\epsilon_{\pm}(k)$ on the edge. The linear bands become two quadratic bands with a gap at $k=0$. In terms of the operators $\psi_{k\pm}$, the singlet pairing term $H_S$ can be written as,
\begin{eqnarray}
H_{S}&=&\sum_{\textbf{k}}[\frac{\Delta_p(\textbf{k})}{2}(\psi_{-\textbf{k}+}\psi_{\textbf{k}+}
+\psi_{-\textbf{k}-}\psi_{\textbf{k}-}+h.c.) \nonumber \\
&&+\Delta_s(\textbf{k})(\psi_{-\textbf{k}+}\psi_{\textbf{k}-}+h.c.)]
\end{eqnarray}
where $\Delta_p(\textbf{k})=-\frac{vk\Delta}{\sqrt{v^2 k^2+h^2}}$ and $\Delta_s(\textbf{k})=-\frac{h\Delta}{\sqrt{v^2 k^2+h^2}}$.   Therefore, effectively  the model describes a system with interband s-wave pairing and intraband $p$-wave pairing. If the Fermi level just crosses one edge band, we can omit the other band and the system is 1D $p$-wave superconducting\cite{Alicea2012}. Provided $h<\sqrt{\Delta^2+\mu^2}$,  the edge forms a topological superconductor\cite{Alicea2012}.  Especially when $h$ vanishes, the system is a gapped topological superconductor with time reversal symmetry.

\begin{figure}[tb]
\centerline{\includegraphics[height=2.7 cm]{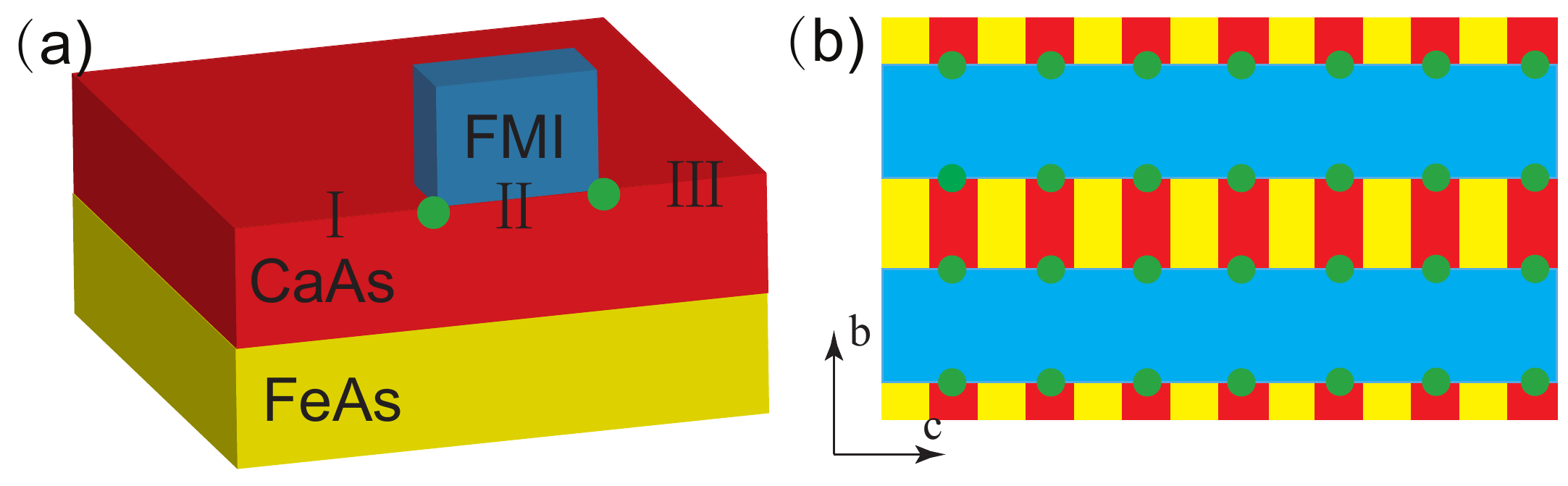}}
\caption{(color online). Schematic picture of the proposed setup for the realization of Majorana fermions. (a) Two Majorana modes. (b) A Majorana chain realization. The green circles denote Majorana modes and the blue rectangles denote ferromagnetic insulator(FMI).
 \label{TS} }
\end{figure}

The above results suggest that the CaFeAs$_2$ is an ideal material to generate and manipulate Majorana modes. Here we discuss two  possible examples. Majorana modes can be trapped in a suitable geometry as shown Fig.\ref{TS}(a)  in which a ferromagnetic insulator is used  and the edge \uppercase\expandafter{\romannumeral2} becomes a normal insulator because of the strong Zeeman field via a proximity effect with  the  ferromagnetic insulator. In this case, experimentally, a quantized $2e^2/h$ zero-bias conductance, a signature for the Majorana modes, should be detected in the topological superconductor with Majorana modes and spinless metal junctions\cite{Alicea2012,Law2009}.  A more interesting case is to consider the [100] surface.  In this case, each CaAs layer has an edge state. Thus, they form a weakly-coupled quasi one-dimensional topological superconductors. By introducing insulating magnetic thin films, as shown in Fig.\ref{TS}(b),  we can design Majorana chains along $c$ direction. This system  carries intriguing physical properties. For example, the neutral Majorana Fermions do not contribute to the electric conductivity but to the thermal conductivity and specific heat. The low-temperature thermal conductivity $\kappa$ of the Majorana Fermions chain is proportional to $T$\cite{Neupert2010}.  As CaFeAs$2$ is most likely to be a full gapped superconductor, the existence of Majorana Fermions can be confirmed by a $T$-linear contribution to the thermal conductivity at low temperature.

The major barrier to observe above predicted physics lies on  the material quality.  Currently,  the quality of the synthesized materials  CaFeAs$_2$  is rather poor as shown in the recent  angle resolved photoemission spectra(ARPES) measurement\cite{Liucpl2013}. It is essential to have high quality single crystal sample with electron doping which can reveal the Dirac point position of the edge states inside the bulk valence band(Fig.\ref{wannier_edge}(c)). In order to use the down edge bands to realize topological superconductors, we also need electron doping to make the chemical potential to cross only the down edge bands.  To realize Majorana Fermions, the candidates of ferromagnetic insulator are CdCr$_2$Se$_4$ and CdCr$_2$S$_4$, which can be deposited on the surface\cite{Baltzer1966}.  Nevertheless,  since   $T_c$  in  Ca$_x$La$_{1-x}$FeAs$_2$ can reach to 45 K\cite{Katayama2013} and the gap caused by the SOC is around 0.1 eV calculated here,  the topological properties are expected to be robust and should be relatively easy to observe  locally within a small clean region even if the overall quality of  a material sample is bad. We suggest that  local probes such as Scanning Tunneling Microscopy (STM) should be deployed to observe the predicted physics first.

Our study suggests a new direction to search QSH insulator materials and their integrations with superconductivity.  From our calculation, it is clear that a separated single As layer, if it can maintain the same structure,  must be a QHS insulator.  In fact,  the crystal structure of recent discovered Phosphorene\cite{Li2014,Liu2014} is similar to that of the As layer although the band gap in Phosphorene is too large to overcome by  the SOC.  The family of iron-based superconductors which include  diversified classes of materials provide  a great opportunity  to search, design and grow new heterostructures  or new  bulk materials for exploring topologically related novel physics together with SC.  In fact,  for heterostructure, one of us\cite{Hao2014} has shown that the  single layer FeSe\cite{Wang2012-fese,Liu2012-fese} grown on SrTiO$_3$ substrates by molecular beam epitaxy (MBE) can be  a QSH insulator without electron doping.

In conclusion, we predict that CaFeAs$_2$ is a staggered intercalation compound that integrates both QSH and SC and is an ideal system for the realization of Majorana Modes.

We thank discussion with H. Ding, T. Xiang, K. K. Li, J. J. Zhou and S. M. Nie.  The work is supported by the Ministry of Science and Technology of China 973
program(Grant No. 2012CV821400 and No. 2010CB922904), National Science Foundation of China (Grant No. NSFC-1190024, 11175248 and 11104339), and   the Strategic Priority Research Program of  CAS (Grant No. XDB07000000).


\end{bibunit}

\begin{bibunit}

\clearpage
\pagebreak
\onecolumngrid
\widetext
\begin{center}
\textbf{\large Supplementary material for ``CaFeAs$_2$: a Staggered Intercalation of Quantum Spin Hall and High Temperature Superconductivity''}
\end{center}


\setcounter{equation}{0}
\setcounter{figure}{0}
\setcounter{table}{0}
\makeatletter
\renewcommand{\theequation}{S\arabic{equation}}
\renewcommand{\thefigure}{S\arabic{figure}}
\renewcommand{\bibnumfmt}[1]{[S#1]}
\renewcommand{\citenumfont}[1]{S#1}

\begin{figure}[bt]
\centerline{\includegraphics[height=6.5 cm]{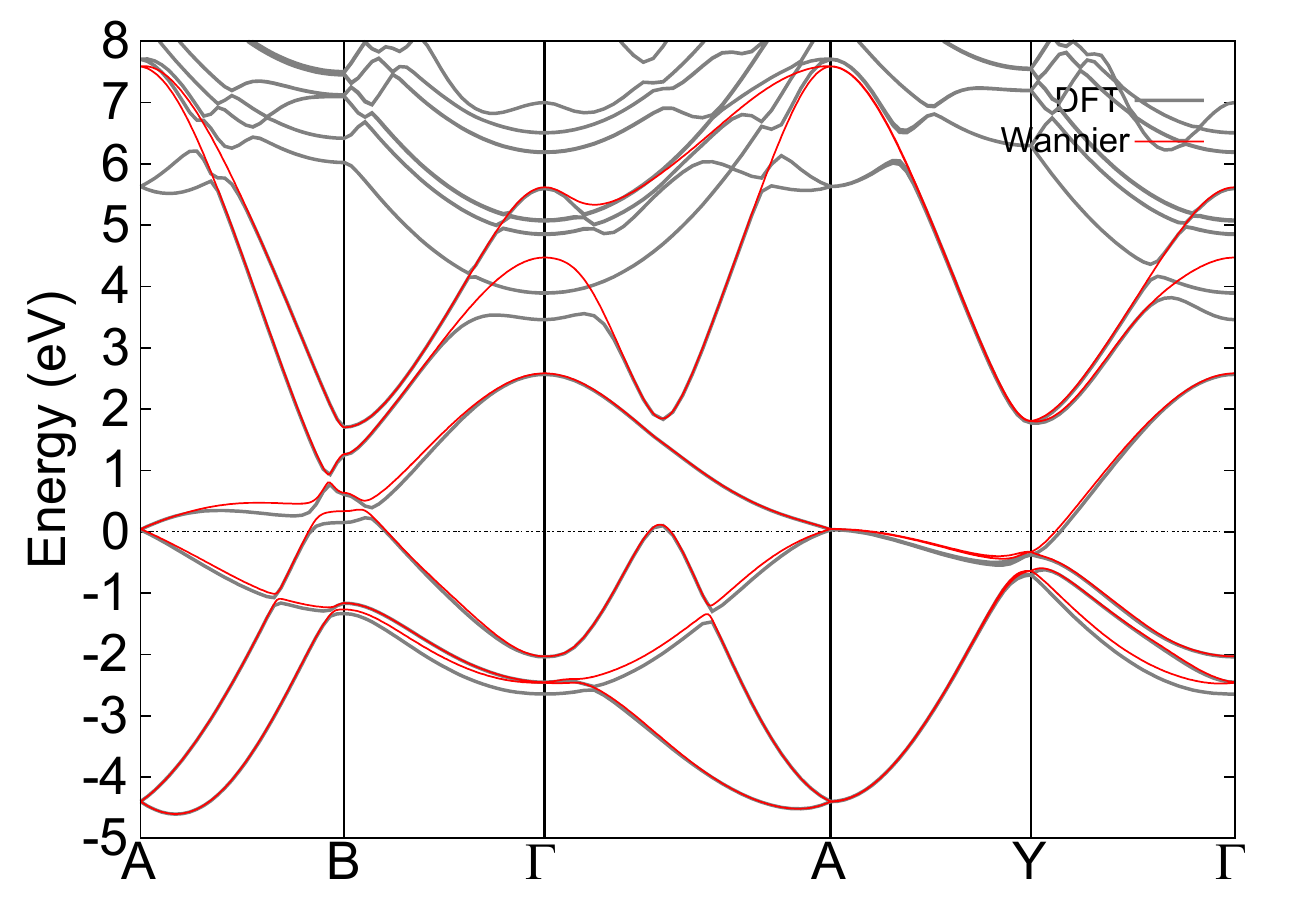}}
\caption{(color online). The band structures of DFT and tight binding model using maximum localized Wannier functions for the free standing As layers with SOC. The gray line represents the DFT band structure and the red line represents the tight-binding band structure.
 \label{dft_tb_As} }
\end{figure}

\begin{figure}[bt]
\centerline{\includegraphics[height=7 cm]{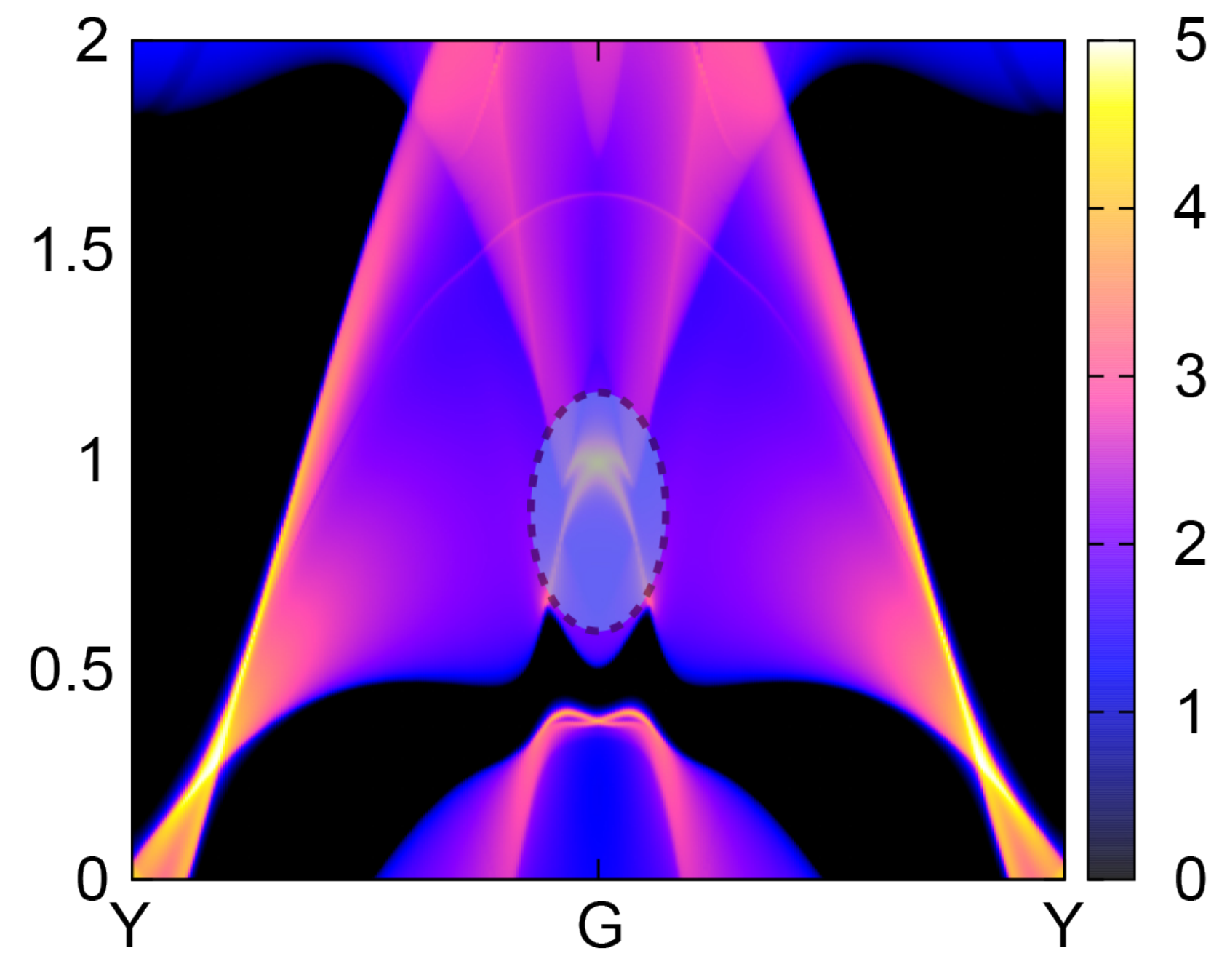}}
\caption{(color online). Energy and momentum dependence of the LDOS for the free standing As layers on the [100] surface. The higher LDOS is represented by brighter color. The nontrivial edge states are around the $\Gamma$ point and highlighted with the ellipse.
 \label{wannier_edge_As} }
\end{figure}

We perform LDA calculation for the free-standing As layer, where the model contains only the As layers and a large vacuum space along c axis is adopted to avoid the interactions between the As layers. The LDA band structure is shown in Fig.\ref{dft_tb_As}(gray lines). The bands of the As layers contributed by $p_x$ and $p_y$ orbitals are quite similar to those of As layers CaFeAs$_2$ in Fig.2(d)). But, the bands attributed to As $p_z$ orbitals show a difference from those of  As layers in CaFeAs$_2$. This is due to the coupling of As1 pz orbitals and Ca d orbitals in CaFeAs$_2$. However, this difference doesn't affect the topology of As layers because the topology is determined by $p_x$ and $p_y$ orbitals. We fit the bands of As layers to a tight-binding Hamiltonian using maximally localized Wannier functions. The tight-binding band is shown in Fig.\ref{dft_tb_As}(red lines). The LDA band and tight-binding band match well, indicating that the tight-binding model is valid. We have calculate the Wannier function centers of the 12 occupied bands and find that the Wannier states switch partners. Thus, the $Z_2$ topological invariant quantum number is odd.  With the above tight-binding model, we can obtain the surface Green's function of the semi-infinite system using an iterative method\cite{Sancho1984,Sancho1985}. The local density of states (LDOS) are just the imaginary parts of the surface Green's function and we can obtain the dispersion of the surface states from the LDOS. The  LDOS of [100] edge for the As layers is shown in Fig.\ref{wannier_edge_As}. The nontrivial edge states around $\Gamma$ point, similar to those in Fig.3(c), show the nontrivial topology of the As layers.  This result proves that  the main conclusions in this paper and further confirms that the As layers is relatively inert.  For the bands near Fermi surfaces, there is very weak coupling between the FeAs layers and CaAs layers.

\begin{figure*}[bt]
\centerline{\includegraphics[height=8 cm]{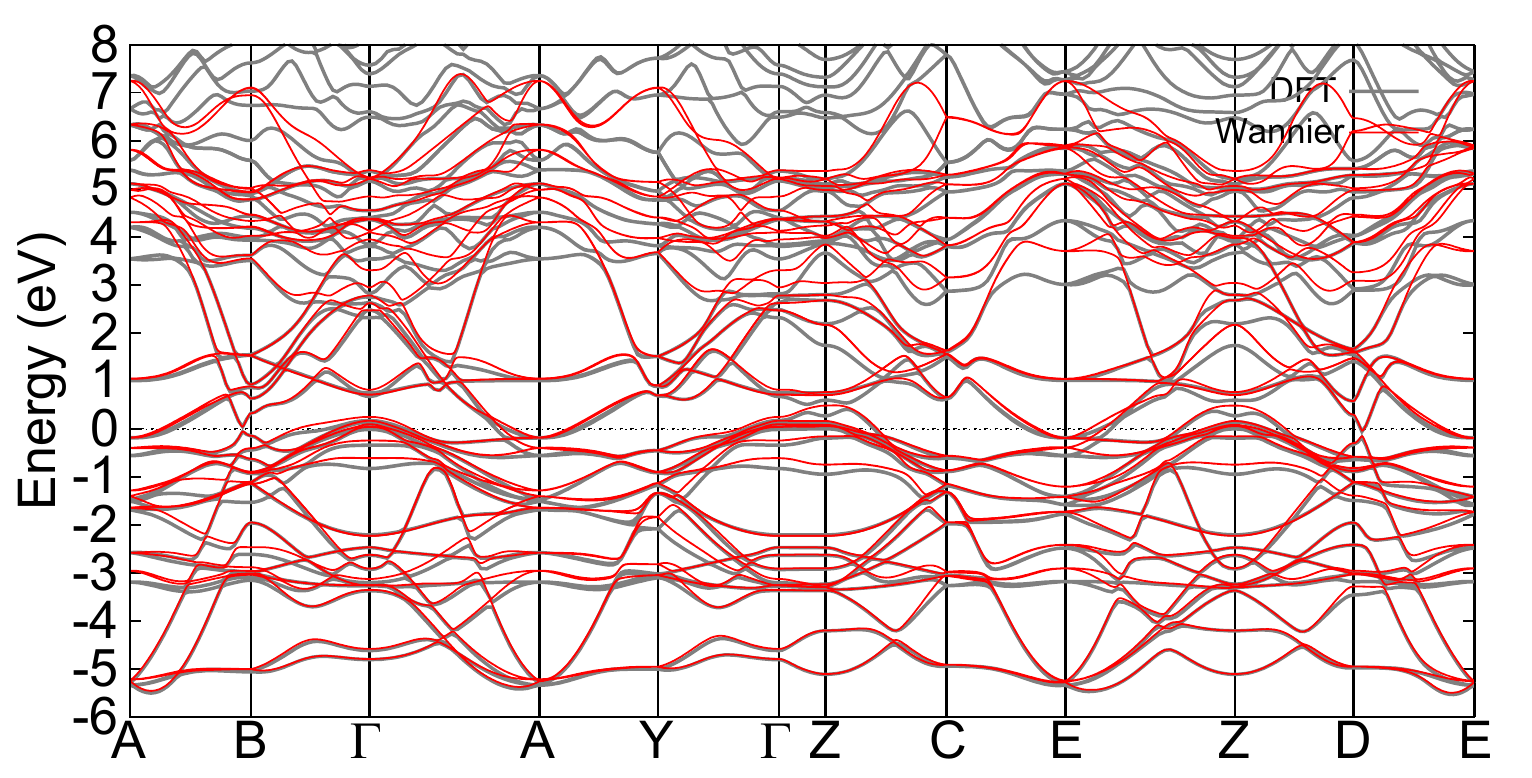}}
\caption{(color online). The band structures of DFT and  the tight binding model using maximum localized Wannier functions for CaFeAs$_2$ with SOC. The gray line represents the DFT band structure and the red line represents the tight-binding band structure.
 \label{dft_tb_CaFeAs2} }
\end{figure*}

In the main text, the edge states are calculated using a tight-binding model with only $p_x$ and $p_y$ orbitals of As-1. However, in a full system,  there could be other trivial edge states from FeAs layers which may couple with the nontrivial edge states of the As layers. It may affect the dispersion of the nontrivial edge states. In the following, we report a calculation based on a full tight-binding Hamiltonian to show that the topological edge states are not affected by other nontrivial edges. We first fit the bands of CaFeAs$_2$ to a tight-binding Hamiltonian using maximally localized Wannier functions, where Fe $3d$, As $4p$ and Ca $4s$, $3d$ orbitals are included. The band structures of DFT and the tight binding model are shown in Fig.\ref{dft_tb_CaFeAs2} and they match well in a wide energy range. The surface LDOS of [100] surface for CaFeAs$_2$ are shown in Fig.\ref{wannier_edge_As}. There are trivial surface states from the FeAs layers. The surface states in the shaded ellipse, similar to those in Fig.\ref{wannier_edge_As}and Fig.3(c), are attributed to the As layers. Thus, the edge states of FeAs layers have little effect on the nontrivial edge states of As layers.

To check the topological properties of the complete CaFeAs$_2$, we also calculated the $Z_2$ topological number. The $Z_2$ topological invariance for a three dimensional system is specified by four indices: the strong topological number $\nu_0$ and three weak topological numbers $\nu_1$, $\nu_2$ and $\nu_3$\cite{Fu2007} as $Z_2=\nu_0;(\nu_1\nu_2\nu_3)$. We calculated the parities of bands at time reversal invariant momenta for an equivalent CaFeAs$_2$ structure with inversion symmetry.  The calculated $Z_2$ invariant is $Z_2=0;(001)$, indicating that CaFeAs$_2$ is a weak topological insulator.

\begin{figure}[bt]
\centerline{\includegraphics[height=7 cm]{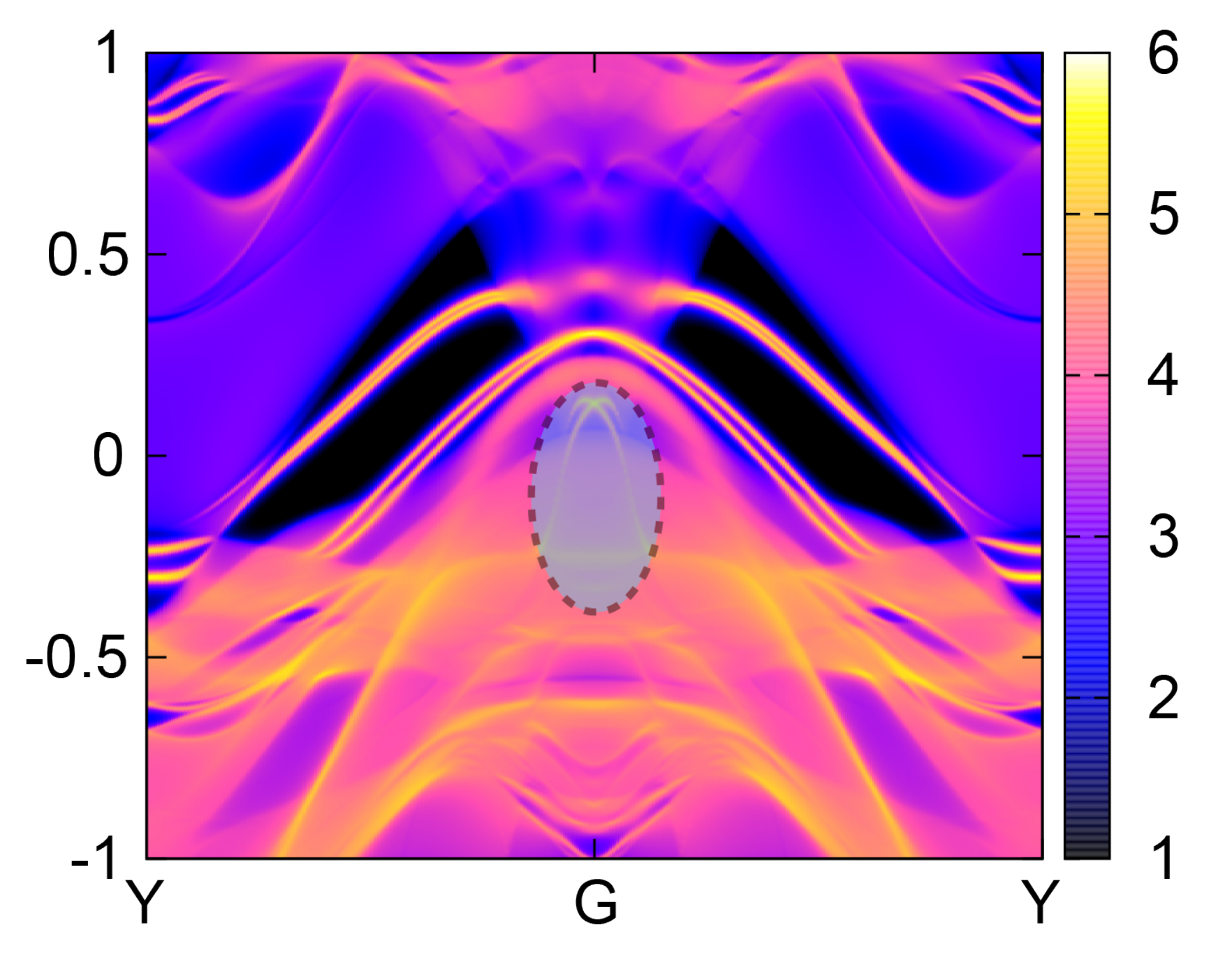}}
\caption{(color online). Energy and momentum dependence of the LDOS for   the [100] surface  CaFeAs$_2$. The higher LDOS is represented by brighter color. The nontrivial edge states, attributed to the $p_x$ and $p_y$ orbitals of As-1, are around the $\Gamma$ point.
 \label{wannier_edge_CaFeAs2} }
\end{figure}


\end{bibunit}

\end{document}